\documentclass[aps,prd,preprint,nofootinbib,superscriptaddress]{revtex4-2}

\usepackage{amsmath,amssymb}
\usepackage{graphicx}
\usepackage[colorlinks=true,allcolors=blue]{hyperref}

\graphicspath{{figures/}}

\newcommand{\wD}{\omega_{\scriptscriptstyle DDE}}
\newcommand{\rD}{\rho_{\scriptscriptstyle DDE}}
\newcommand{\rM}{\rho_{\scriptscriptstyle M}}
\newcommand{\rN}{\rho_{\scriptscriptstyle NDE}}
\newcommand{\chiME}{\chi_{\scriptscriptstyle ME}}

\begin{document}

\title{Matter-Dark Energy Transition: A Dynamical Systems Approach}

\author{Bob Osano}
\email{bob.osano@uct.ac.za}
\affiliation{Cosmology and Gravity Group, Department of Mathematics and Applied Mathematics, University of Cape Town (UCT), Rondebosch 7701, Cape Town, South Africa}
\affiliation{Centre for Higher Education Development, University of Cape Town (UCT), Rondebosch 7701, Cape Town, South Africa}

\date{November 4, 2024}

\begin{abstract}
In this study, we deviate from the traditional examination of the evolving universe through the scale factor and instead consider a parameter $\chiME$, defined by the ratio of two competing energy densities: matter (DM) and dynamic dark energy (DDE). While the scale factor's role is not neglected, it is inherently embedded within the evolution equations of these competing energy densities. By employing dynamical systems techniques, we investigate the behaviour of this ratio $\chiME$ to understand the dynamics surrounding the cosmological transition from matter domination to dark energy domination. This methodological shift allows for a more nuanced analysis of the matter-to-dark energy transition.
\end{abstract}


\maketitle

\section{Introduction}

Despite significant progress in cosmology, our comprehension of dark energy remains incomplete. A range of dark energy models, especially those beyond the scope of the cosmological constant, have been proposed to explain its characteristics. While some models align partially with observational data, this alignment typically occurs only under specific conditions or parameter settings. Such limitations underscore the necessity for alternative frameworks, possibly beyond conventional field theories, to probe innovative possibilities for dark energy. These developments open the door to examining scenarios where the dark energy equation of state (EoS) deviates from the classical assumption, $\omega_{DE}\neq-1$, potentially offering new insights into its properties and behaviour.

Dynamic dark energy (DDE) models, in particular, have been extensively investigated in the literature~\cite{Planck2018,Zhao2017,Yang2019,DiValentino2021,Vagnozzi2020,Kelley2019,Joudaki2017,DiValentino2017,Yang2021,Benisty2022,Roy2024,DESI2024,Sharma2022}. In many cases, the effective $\omega_{DE}$ is reconstructed from observational data, exemplified in~\cite{Zhao2017}. These studies often favour DDE models featuring a phantom-like EoS ($\wD<-1$). Nonetheless, it has been demonstrated that such models provide only partial resolutions to the ongoing $H_0$ tension. However, these models possess certain limitations. Specifically, in some instances, they exacerbate the growth tension associated with Cold Dark Matter (CDM) models and provide an inferior fit to Baryon Acoustic Oscillation (BAO) and Type~Ia Supernova (SNIa) data compared to CDM~\cite{Alestas2021}. Additionally, they tend to slightly favour an absolute magnitude value for SNIa that is only marginally consistent with measurements derived from Cepheid variable stars~\cite{Alestas2021}. Despite these shortcomings, it is crucial to acknowledge that these models have not been definitively ruled out. Research has demonstrated~\cite{DiValentino2021e} that certain DDE models with a phantom-like EoS can potentially address the Hubble constant tension when analysed alongside BAO and Pantheon data. Moreover, DDE models appear capable of alleviating the $\sigma_8$ tension~\cite{Chudaykin2021} to some extent. A comprehensive summary of various tensions in the field of cosmology can be found in~\cite{Abdalla2022}.

These findings suggest that a Dynamic Dark Energy (DDE) model that aligns with observational data may still be a viable alternative. In this study, we utilize dynamical systems techniques to investigate a generic DDE model within the framework of a cosmological transition from matter domination to dark energy (DE) domination. Given the distinct evolutionary dynamics of different energy-matter densities, we aim to uncover what insights can be gleaned about the equation of state parameter $\wD$ through an analysis of a flat $\Lambda$CDM model. This investigation examines the parameter $\chiME$, which denotes the ratio of DDE to matter density (DM). The results of this analysis are presented in this article.

\section{Cosmological model}

The success of the $\Lambda$CDM model, or the standard model of Big Bang cosmology, in reconciling theory with a multitude of observations makes it a highly suitable starting point for any cosmological study. This model, employing general relativity as the theory of gravity on cosmological scales, successfully accounts for multiple key observations: the cosmic microwave background, the distribution of galaxies, the abundances of light elements, and cosmic acceleration as evidenced by light from distant galaxies and supernovae. Despite these achievements, we remain cognizant of its limitations. These limitations are precisely what spur contemporary research in the field of cosmology.

Regarding cosmic acceleration, the prevailing hypothesis suggests that an unknown form of energy density is responsible for this phenomenon. Consequently, the various studies highlighted in the previous section explore models with different formulations of the equation of state (EoS). In our analysis, we employ a generic $\wD$ and utilize dynamical systems techniques to identify structural changes in the flow, thereby linking these changes to alterations in the evolution of the cosmological model.

\subsection{Friedmann equations}

We consider the Lambda Cold Dark Matter ($\Lambda$CDM) model and employ the Friedmann equations as the foundation of our discussion. In cosmic time, these equations are expressed as follows:
\begin{align}
\frac{\dot a^2}{a^2} &= \frac{8\pi G}{3}\rho_\Sigma-\frac{k}{a^2}+\frac{\Lambda}{3}, \label{eq:F1}\\
\frac{\ddot a}{a} &= -\frac{4\pi G}{3}\left(\rho_\Sigma+3p_\Sigma\right)+\frac{\Lambda}{3}, \label{eq:F2}
\end{align}
where the overdot denotes a derivative with respect to cosmic time $t$. Here, $a(t)$ represents the scale factor, $G$ is the Newtonian gravitational constant, $\rho_\Sigma$ is the cumulative density of all components in the model, $p_\Sigma$ is the total pressure, and $\Lambda$ is the cosmological constant. In equations~\eqref{eq:F1} and~\eqref{eq:F2}, we have set the speed of light to 1 ($c=1$) and will later set $8\pi G=1$.

By utilizing the logarithmic scale $\eta=\log a$ for time, where a derivative with respect to $\eta$ is denoted by a prime ($'$), the equations can be reformulated as follows:
\begin{align}
\left(\frac{a'}{a}\right)^2 &= \frac{8\pi G}{3H^2}\rho_\Sigma-\frac{k}{a^2H^2}+\frac{\Lambda}{3H^2}, \label{eq:F1eta}\\
\frac{\ddot a}{aH^2} &= 1+\frac{H'}{H} = -\frac{4\pi G}{3H^2}\left(\rho_\Sigma+3p_\Sigma\right)+\frac{\Lambda}{3H^2}, \label{eq:F2eta}
\end{align}
where $a(t)$ is the scale factor, $G$ is the Newtonian gravitational constant, $\rho_\Sigma$ is the cumulative density of all the components of the model, $p_\Sigma$ represents the cumulative pressure, and $\Lambda$ is the cosmological constant. Here, $k$ is the spatial curvature and $H$ is the Hubble parameter. Setting $\eta=\log a$ simplifies the expressions and provides a useful framework for analysing the dynamical evolution of the cosmological scale factor.

Equations~\eqref{eq:F1eta} and~\eqref{eq:F2eta}, which are our main constraint equations, are simply
\begin{align}
1 &= \frac{8\pi G}{3H^2}\rho_\Sigma-\frac{k}{a^2H^2}+\frac{\Lambda}{3H^2}, \label{eq:C1}\\
\frac{H'}{H} &= -1-\frac{4\pi G}{3H^2}\left(\rho_\Sigma+3p_\Sigma\right)+\frac{\Lambda}{3H^2}, \label{eq:C2}
\end{align}
since $a'/a=1$ and $\ddot a/(aH^2)=1+H'/H$. The energy density $\rho_\Sigma$ is made up of radiation $\rho_r$, ordinary matter $\rho_m$, dark matter $\rho_{DM}$, and dynamic dark energy $\rD$. We can think of the cosmological constant as a form of energy density capable of driving expansion. We can represent this using a non-dynamic energy density denoted $\rN$ which has the form $\rN=\Lambda/(8\pi G)$. In this case equation~\eqref{eq:F1eta} takes the form:
\begin{equation}
1=\frac{8\pi G}{3H^2}\left(\rho_\Sigma+\rN\right)-\frac{k}{a^2H^2}. \label{eq:C1NDE}
\end{equation}

To further establish rigour in our formulation, we incorporate the constant $8\pi G$ explicitly into our equations, a step we had deferred until defining $\rN$. In equation~\eqref{eq:C1NDE}, we posited that the non-dynamic dark energy component is solely made up of the cosmological constant, $\Lambda$. This assumption is not dictated by any fundamental principle but rather by our intent to simplify the model by minimising the number of variables.

It's important to recognise that the equation of state for dark energy, which we denote as $\omega_{DE}$, can be viewed as an effective parameter that encapsulates contributions from both dynamic dark energy ($\wD$) and non-dynamic dark energy ($\omega_{NDE}$). This composite state can be expressed as a linear combination:
\begin{equation*}
\omega_{DE}=C_1\wD+C_2\,\omega_{NDE},
\end{equation*}
where $C_1=\rD/\rho_{DE}$ and $C_2=\rN/\rho_{DE}$. These coefficients, which in general vary with time, are determined based on the proportion of each individual density relative to the aggregate dark energy density. Specifically, they reflect how much each component---dynamic ($\rD$) and non-dynamic ($\rN$)---contributes to the total dark energy density:
\begin{equation*}
\rho_{DE}=\rD+\rN.
\end{equation*}
By accurately determining these coefficients, $C_1$ and $C_2$, we can capture the influence of each type of dark energy on the overall dynamics of the universe. This approach may provide a more comprehensive understanding of the role dynamical dark energy plays in cosmic evolution.

We introduce hypothetical interactions into our equations of motion. The first interaction considered is between radiation and matter, denoted as $Q_{rm}$. The second interaction is within the dark sector, involving dark matter and dark energy, denoted as $Q_{ME}$ in our framework. For the sake of simplicity and in line with the analysis presented by~\cite{Barkana2018}, we do not account for any interactions between dark matter and baryons.

The evolution equations for the energy densities with respect to the dimensionless time parameter $\eta=\log a$ are given by:
\begin{align}
\rho_r' &= -3(1+\omega_r)\rho_r+\frac{Q_{rm}}{H}, \label{eq:rr}\\
\rho_m' &= -3(1+\omega_m)\rho_m-\frac{Q_{rm}}{H}, \label{eq:rm}\\
\rho_{DM}' &= -3(1+\omega_{DM})\rho_{DM}+\frac{Q_{ME}}{H}, \label{eq:rDM}\\
\rD' &= -3(1+\wD)\rD-\frac{Q_{ME}}{H}, \label{eq:rDDE}\\
\rN' &= -3(1+\omega_{NDE})\rN. \label{eq:rNDE}
\end{align}
These interactions and their corresponding equations are critical for accurately modelling the dynamic evolution of the universe's contents, providing keen insights into the intricate interplay between different forms of matter and energy as the universe evolves. We used the standard definition where the expansion parameter $\Theta=3\dot a/a=3H$, and where the overdot denotes differentiation with respect to cosmic time. A crucial feature of this model is the dichotomy within the dark energy sector: it consists of a portion that interacts with dark matter and a portion that remains non-interacting. The equations of state for all components except for the dynamic dark energy are well-established. Specifically, $\omega_r=1/3$ for radiation, $\omega_m=0$ for matter, $\omega_{DM}=0$ for dark matter, and $\omega_{NDE}=-1$ for non-dynamic dark energy.

To account for the matter density, we denote $\rM=\rho_m+\rho_{DM}$, where $\rho_m$ is the density of baryonic matter and $\rho_{DM}$ is the density of dark matter. For dark energy, we use $\rho_{DE}=\rD+\rN$, where $\rD$ represents the density of dynamic dark energy and $\rN$ represents the density of non-dynamic dark energy. These expressions will be utilised to incorporate the varying densities across different epochs in the universe's evolution.

\subsection{Transition dynamics}

Although the method developed in this study can be adapted to examine cosmological dynamics at any epoch, we specifically apply it to investigate the dynamics surrounding the transition from matter domination to dark energy domination. While multiple transitional periods exist in cosmology, our focus is restricted to the critical transition where the dominant driving force of the cosmological expansion shifts from matter to dark energy. This period is pivotal as it signifies a shift in the primary component influencing the universe's expansion.

The transition is characterised by a transfer of dominance from one form of energy density to another and can be intuitively understood as a change in the primary driver of cosmological expansion. These transitions are feasible because the different constituent energy densities evolve according to distinct equations. This differentiation is well captured in the evolution equations, as indicated by equations~\eqref{eq:rr}--\eqref{eq:rNDE}. 
Essential to their evolution is their respective equation of state, which dictates how each form of energy density changes over time.

By examining this transition, we aim to gain deeper insights into the dynamics of cosmological expansion, particularly in the context of how varying components contribute differently over time. This approach not only enriches our understanding of the individual contributions of matter and dark energy but also informs our broader comprehension of the universe's expansion history.

\subsection{Matter-to-dark energy transition}

To embark on a detailed examination of a flat $\Lambda$CDM model, let us also note that recent discrepancies in cosmological observation data~\cite{Abdalla2022} call for a reassessment of the standard flat $\Lambda$CDM paradigm. Incorporating the transition effect from a matter-dominated universe to one dominated by dark energy into the Friedmann equations is essential. By abstracting the matter density $\rM$, we obtain a simplified form of the Friedmann equations. This is achieved by using Equation~\eqref{eq:C1} to eliminate $\Lambda$ from Equation~\eqref{eq:C2}, with $k=0$, $8\pi G=1$ and radiation neglected, yielding the following resultant equation:
\begin{equation}
\frac{H'}{H}=-\frac{\rM}{2H^2}\left[1+(1+\wD)\frac{\rD}{\rM}\right], \label{eq:Hprime}
\end{equation}
since $\omega_M=0$. Note that $\rN$, equivalently $\Lambda$, identically cancels in the subtraction process and does not feature in equation~\eqref{eq:Hprime}. Subsequent derivations and analysis should take into account both interacting and non-interacting components of dark energy, as discussed previously. This nuanced approach may provide fresh insights into understanding the tensions arising from observational data and contribute to refining our cosmological models.

Equation~\eqref{eq:Hprime} can be interpreted in relation to the deceleration parameter, $q$, as follows:
\begin{equation}
\frac{H'}{H}=\frac{\dot H}{H^2}=-(1+q). \label{eq:q}
\end{equation}
Deceleration corresponds to $q>0$ and acceleration to $q<0$, with $q=-1$ for de~Sitter expansion. In the far future of a $\Lambda$CDM model $q$ tends to $-1$ from above and the Hubble parameter asymptotes to the constant value $\sqrt{\Lambda/3}=H_0\sqrt{\Omega_\Lambda}$, where $H_0$ is the Hubble constant today. Models of the universe filled with non-phantom energy other than $\Lambda$ decelerate in the intermediate regime. This implies a decelerating universe for any mass-energy density with EoS $\omega>-1/3$. Recent observations of distant type~Ia supernovae~\cite{Turner2000} indicate that $q$ is negative or that the expansion of the universe is accelerating. This intimates that some form of energy, dark energy, is counteracting the gravitational pull of matter, on the cosmological scale. It is the transition to this domination that is of interest to us. We introduce a new dimensionless parameter that will allow equation~\eqref{eq:Hprime} to be expressed in terms of two competing energy densities at transition. We let
\begin{equation}
\chiME=\frac{\rD}{\rM}-1. \label{eq:chi}
\end{equation}
In this paper, we employ techniques from nonlinear dynamical systems theory. Instead of explicitly determining solutions to a system of equations, our primary interest lies in understanding the behavioural changes in the system's flow. Nonlinear dynamical systems techniques provide the necessary tools for this analysis. Consequently, we will briefly review some key concepts in nonlinear dynamical systems that are relevant to our study. This field is both broad and deep, so the following presentation is only a brief overview. Readers interested in a more comprehensive understanding may consult the extensive range of books and resources available on the techniques and applications of nonlinear dynamical systems.

\section{Dynamical systems technique in cosmology}

To contextualise our research, we first discuss the applicability of the dynamical systems approach in cosmology. It is well-understood that the evolution of systems in continuous time can be characterised using differential equations. In the cosmological context, non-linear dynamical systems techniques allow us to recast a system of differential equations governing cosmological evolution into a form that manifests the fundamental properties of the model under examination. 
The key feature of this technique, when compared to numerical or analytical techniques~\cite{Osano2024a}, is that our focus is on the overall dynamics rather than individual solutions. In this regard, the dynamical systems approach enables us to gain a comprehensive understanding of how different cosmological parameters interact and evolve over time~\cite{Bahamonde2018}.

Our study is about cosmological transitions and this approach is particularly valuable for modelling transitions between different phases of matter in the universe, such as from radiation-dominated to matter-dominated eras. Utilising Friedmann's equations as the foundation, supplemented by principles from particle physics, we can construct a system of equations that accurately describes these transitional periods. As we will demonstrate, dynamical systems techniques enable the analysis of this system through phase space diagrams, stability analysis, and the identification of fixed points, which correspond to important points in the evolution of the universe.

The stability analysis of the fixed points allows us to determine potential cosmic attractors, and therefore comment on the future state of the universe. In our context, cosmic attractors are the stable configurations or evolutions of the universe's expansion. They are specific solutions or states in cosmological models that represent long-term, stable behaviours towards which the universe naturally evolves over time. They represent the equilibrium state of the evolution. They will manifest as points, curves, or more complex structures in phase space where trajectories from a variety of initial conditions converge. Several cosmic attractors in cosmology are already known. These attractors which occur in different phases of evolution include:
\begin{itemize}
\item Slow-roll inflation attractors~\cite{Pattison2018}: In early universe cosmology, during inflation, the slow-roll conditions for the inflaton field often act as attractors. As the inflaton field slowly evolves down its potential, the inflationary dynamics drive the universe toward a rapid expansion phase, which stabilizes the universe's early state.
\item Scaling solutions~\cite{Copeland1998}: In some cosmological models, especially those involving scalar fields (like quintessence), the field's energy density can scale with the matter or radiation energy density. These scaling solutions can act as attractors, where the scalar field mimics the evolution of other components of the universe over time.
\item Matter-dominated attractors~\cite{Papagiannopoulos2020}: The universe evolves towards a state dominated by matter, typically occurring after radiation domination and before dark energy becomes significant.
\item Dark energy-dominated attractors~\cite{ArmendarizPicon2000}: The universe evolves towards an accelerated expansion phase, dominated by dark energy, which may correspond to the current and future state of the universe.
\end{itemize}
The classification of cosmic attractors as stable or unstable fundamentally hinges on the system's response to small perturbations or variations in initial conditions. A stable cosmic attractor is characterised by the system's tendency to evolve toward predictable, self-similar behaviour, for example, the accelerating expansion of the universe, often described in models incorporating a cosmological constant or a scalar field. Conversely, unstable cosmic attractors, although they do not denote final states, are crucial for understanding the dynamics of transitional periods in cosmological evolution, offering insights into the mechanisms and processes that shape the universe's trajectory. Our study differs from most previous studies and focuses on a potentially unstable attractor marking the transition to a dark-energy-dominated universe.

We consider a dynamical system of any number of equations with the form:
\begin{equation}
x'^{(n)}(\eta)=f^{(n)}\!\left(x^{(m)}(\eta)\right), \label{eq:gensys}
\end{equation}
where the prime above a variable denotes its derivative with respect to time $\eta$, and $f$ represents a function often composed of linear or nonlinear terms related to the variables $x^{(m)}$ and their interactions. A variety of functions $f$ typically present themselves in a nonlinear form. The transition dynamics discussed in this article illustrate one such instance. A critical point for the system satisfies $f(x^{*(m)})=0$. We then linearise the system around the fixed points $x^{*(m)}$ which we express as follows:
\begin{equation}
\delta x'^{(n)}=J^{(n)}{}_{(m)}\,\delta x^{(m)},\qquad
J^{(n)}{}_{(m)}=\left.\frac{\partial f^{(n)}(x^{(m)})}{\partial x^{(m)}}\right|_{x^{(m)}=x^{*(m)}}, \label{eq:lin}
\end{equation}
where $J^{(n)}{}_{(m)}$ is the respective Jacobian. It is straightforward to determine the eigenvalue $\lambda_j$ which corresponds to the individual fixed point. The corresponding eigenfunctions are proportional to $e^{\lambda_j\eta}$. We say $x^{*(m)}$ is attracting if there exists a $\delta>0$ such that
\begin{equation*}
\lim_{\eta\to\infty}x^{(m)}(\eta)=x^{*(m)}\quad\text{whenever}\quad\|x^{(m)}(0)-x^{*(m)}\|<\delta.
\end{equation*}
It is important to define a Lyapunov stable point as well because the interpretation in terms of cosmological attraction will have a profound consequence. 
We say a fixed point $x^{*(m)}$ is Lyapunov stable if for each $\epsilon>0$, there exists a $\delta>0$ such that
\begin{equation*}
\|x^{(m)}(\eta)-x^{*(m)}\|<\epsilon\quad\text{for all }\eta\geq0\quad\text{whenever}\quad\|x^{(m)}(0)-x^{*(m)}\|<\delta.
\end{equation*}
In other words, the trajectories that start within $\delta$ of $x^{*(m)}$ remain within $\epsilon$ of $x^{*(m)}$ for all positive $\eta$. In the context of cosmological transition dynamics, a Lyapunov stable and attracting fixed point---meaning it is asymptotically stable---indicates a final state where the system eventually settles. However, the specific nature of this transition is significantly influenced by interaction terms. These terms can cause the trajectories to initially resist stabilisation, delaying their convergence to the attractor. Let us first formulate the system of equations that characterises dynamics around the matter-dark energy transition.

\section{Matter to dynamic dark energy transition}

To simplify the notation, we will relabel $\chiME$ in equation~\eqref{eq:chi} as $X$. In addition to this, we introduce three other dimensionless variables that emerge when $\chiME$ is differentiated with respect to $\eta$. The complete set of variables is defined as:
\begin{equation}
X=\frac{\rD}{\rM}-1,\qquad Y=\frac{Q_{ME}}{\rM H},\qquad Z=\frac{\rM}{H^2},\qquad S=\frac{\Lambda}{2H^2}. \label{eq:vars}
\end{equation}

Before presenting the evolution equations for these variables, we need to discuss the effect of interactions on the conditions satisfied by these variables. It is typically acceptable to assume $\rM>0$ and $\rD>0$ in the absence of interaction, which implies that $X\geq-1$. However, when interactions between $\rM$ and $\rD$ are present, as is the case in the system we will examine, assuming $\rM>0$ and $\rD>0$ could lead to the exclusion of potentially viable solutions in the phase space, as demonstrated in~\cite{Quartin2008}. Therefore, while relaxing the positivity conditions might be necessary, it should be approached with caution.

It is suggested that such relaxation may partly explain the difficulty in defining energy density variables in the case of several interacting fluids, as discussed in~\cite{Tamanini2015}. The key takeaway here is that any results obtained must be interpreted within the context of this relaxation condition. In our analysis, not requiring $\rM>0$ and $\rD>0$ alters the previous constraint of $X\geq-1$, allowing $X$ to be lower than $-1$. As we will see, this introduces challenges in interpreting results that fall within the regions where $X<-1$. A discussion of the coupling interaction term would be incomplete without speculating about its potential form. While we do not explicitly propose an ansatz for the interaction term in this study, existing literature already offers some possible forms. For example, $Q\propto\rho_m\rho_{de}/H$ as shown in~\cite{Perez2014}, or $Q\propto\rM X$ as suggested in~\cite{Osano2024a}. Returning to the variables given in equation~\eqref{eq:vars}, it can be demonstrated that differentiating each variable with respect to $\eta$, and making substitutions using equations~\eqref{eq:rDM}--\eqref{eq:rDDE} and variables~\eqref{eq:vars}, results in the following closed system of evolution equations. Since no ansatz for the interaction is adopted, we take $Q_{ME}$ to be constant in $\eta$ ($Q_{ME}'=0$); any other choice modifies only the $Y'$ equation. With
\begin{equation}
D\equiv1+(1+\wD)(X+1),\qquad\text{so that}\qquad \frac{H'}{H}=-\frac{ZD}{2} \label{eq:D}
\end{equation}
by Eq.~\eqref{eq:Hprime}, the system reads
\begin{align}
X' &= -3\wD(X+1)-Y(X+2), \nonumber\\
Y' &= Y\left(3-Y+\frac{ZD}{2}\right), \nonumber\\
Z' &= Z\left(-3+Y+ZD\right), \nonumber\\
S' &= SZD. \label{eq:system}
\end{align}
For a flat universe with radiation neglected, the Friedmann equation~\eqref{eq:C1} becomes the constraint
\begin{equation}
C\equiv\frac{Z}{3}(X+2)+\frac{2}{3}S=1. \label{eq:constraint}
\end{equation}
This constraint is preserved by the flow~\eqref{eq:system}: a direct calculation gives
\begin{equation}
C'=ZD\,(C-1), \label{eq:Cdot}
\end{equation}
so the surface $C=1$ is invariant. Moreover, $S$ does not appear in the first three equations of~\eqref{eq:system}, because $\Lambda$ cancels from Eq.~\eqref{eq:Hprime}. On the physical surface the dynamics is therefore that of the three-dimensional system $(X,Y,Z)$, with $\Lambda$ recovered from
\begin{equation}
S=\frac{3-Z(X+2)}{2}. \label{eq:Sslave}
\end{equation}
Two further invariant sets are useful: the plane $Y=0$ (no interaction) and the plane $Z=0$ ($\rM\ll H^2$). The physically admissible region is $Z\geq0$ and $S\geq0$ ($\Lambda\geq0$), i.e.\ $0\leq Z\leq3/(X+2)$, together with $X\geq-1$ when both $\rM$ and $\rD$ are positive.

The fixed points of the reduced system $(X,Y,Z)$ are:
\begin{align}
P_M &= (-1,\,0,\,3), & S&=0, & \frac{H'}{H}&=-\frac{3}{2}, \label{eq:PM}\\
P_{dS} &= (-1,\,0,\,0), & S&=\frac{3}{2}, & \frac{H'}{H}&=0, \label{eq:PdS}\\
\mathcal{L} &= \left(X_*,\,3,\,Z\right),\ Z\ \text{free}, & S&=\frac{3}{2}-\frac{\wD Z}{2(1+\wD)}, & \frac{H'}{H}&=0, \label{eq:L}
\end{align}
where
\begin{equation}
X_*=-\frac{\wD+2}{\wD+1},\qquad \wD\neq-1. \label{eq:Xstar}
\end{equation}
$P_M$ is the matter-dominated (Einstein--de Sitter) point, with $q=1/2$. $P_{dS}$ is the de~Sitter point, $q=-1$, dominated by $\Lambda$. $\mathcal{L}$ is a line of fixed points on which $D=0$, i.e.\ $\rM+(1+\wD)\rD=0$: the interaction holds the effective equation of state of the matter--DDE fluid at $-1$, so that $q=-1$ along the whole line. Its points include $P_Q=(X_*,3,0)$, where $S=3/2$, and $P_L=(X_*,3,3(1+\wD)/\wD)$, where $S=0$; $P_L$ has $Z\geq0$ only for $\wD<-1$, so for $-1<\wD<0$ the whole half-line $Z\geq0$ of $\mathcal{L}$ has $S>0$. Since $X_*+1=-1/(1+\wD)$, the line lies in the region $X\geq-1$ only for phantom dark energy, $\wD<-1$; for $-1<\wD<0$ it has $X_*<-1$ and hence $\rD<0$. For $\wD=-1.5$, for example, $X_*=1$ ($\rD=2\rM$) and $S\geq0$ requires $0\leq Z\leq1$. For $\wD=-1$ the line is absent.

\section{Analysis and discussion}

The stability of these points is derived in Section~\ref{sec:stability}; we summarise it here. The matter point $P_M$ has eigenvalues $3$, $9/2$ and $-3\wD$, all positive for $\wD<0$: it is the source (past attractor) of the flow, as expected of the matter era. The de~Sitter point $P_{dS}$ has eigenvalues $-3$, $3$ and $-3\wD$ and is a saddle. Within the non-interacting plane $Y=0$, its eigenvalue along $X$ is $-3\wD>0$, so trajectories leave $X=-1$ and $X$ grows without bound: the late-time state is dynamic dark energy domination, $\rD/\rM\to\infty$, which is not a fixed point at finite $X$. The line $\mathcal{L}$ is non-hyperbolic (one zero eigenvalue along the line). Its two other eigenvalues, Eq.~\eqref{eq:eigL}, satisfy $\lambda_-<0$ for $\wD>-2$, while $\lambda_+<0$ if and only if $Z<-6(1+\wD)/\wD$. For phantom DDE ($\wD<-1$) this bound is negative, so $\mathcal{L}$ is of saddle type along its whole physical segment; for $-1<\wD<0$ the segment $0\leq Z<-6(1+\wD)/\wD$ is attracting, but it lies at $X_*<-1$. Consequently there is no finite attractor with $\rM,\rD>0$ for $\wD<0$.

The interaction changes this picture in a way that depends on its sign and on $\wD$. In the plane $Z=0$ the point $P_Q$ has eigenvalues $-3$ and $-3(1+\wD)$: for $\wD>-1$ it attracts trajectories with $Y>0$, but it lies at $X_*<-1$. For phantom DDE, $\wD<-1$, $P_Q$ lies in the physical region and is a saddle. Integrating the full system forward from the present state, Eq.~\eqref{eq:X0}, for $\beta\in\{0,0.5,1\}$ and $|Y_0|\in\{0.05,0.3\}$ shows three outcomes. (i) For $Q_{ME}>0$ and $\wD<-1$, $X\to\infty$ with $Y\to\mathrm{const}$: dynamic dark energy domination. (ii) For $Q_{ME}>0$ and $\wD\geq-1$, energy flowing from DDE to dark matter drives $\rD$ negative; the expansion then either stalls ($Z\to\infty$, i.e.\ $H\to0$, signalling recollapse) or, for weak coupling and $\Lambda>0$, settles on the attracting part of $\mathcal{L}$ at $X_*<-1$. (iii) For $Q_{ME}<0$, the matter density reaches zero at finite scale factor, $a/a_0\approx2$--$5$ ($Y\to-\infty$), for every $\wD$ considered. Outcomes (ii) and (iii) are the instabilities of a constant interaction anticipated in~\cite{Quartin2008,Tamanini2015}: a constant $Q_{ME}$ is acceptable as a description around the transition, but not as a late-time model. They occur in the future and do not affect the past evolution discussed below.

\begin{figure}[t]
\centering
\includegraphics[width=\textwidth]{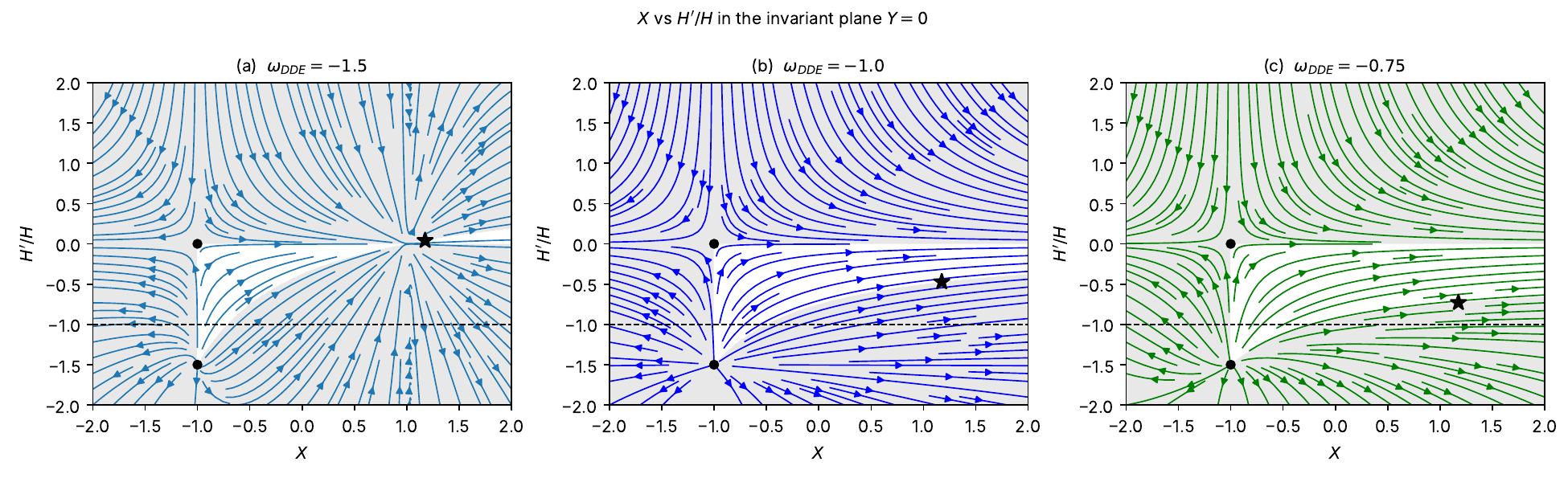}
\caption{Phase portraits $X$ (horizontal axis) and $H'/H$ (vertical axis) in the invariant plane $Y=0$. (a) $\wD=-1.5$, (b) $\wD=-1.0$, (c) $\wD=-0.75$. Dots: $P_M$ at $H'/H=-3/2$ and $P_{dS}$ at $H'/H=0$. Dashed: $q=0$. Star: present state. Grey: excluded region. In (a) the map $Z\mapsto H'/H$ degenerates at $X=1$, where $D=0$ and every trajectory crosses $H'/H=0$; the apparent node at $(1,0)$ is an artefact of the coordinates, not a fixed point.}
\label{fig:XH}
\end{figure}

\begin{figure}[t]
\centering
\includegraphics[width=\textwidth]{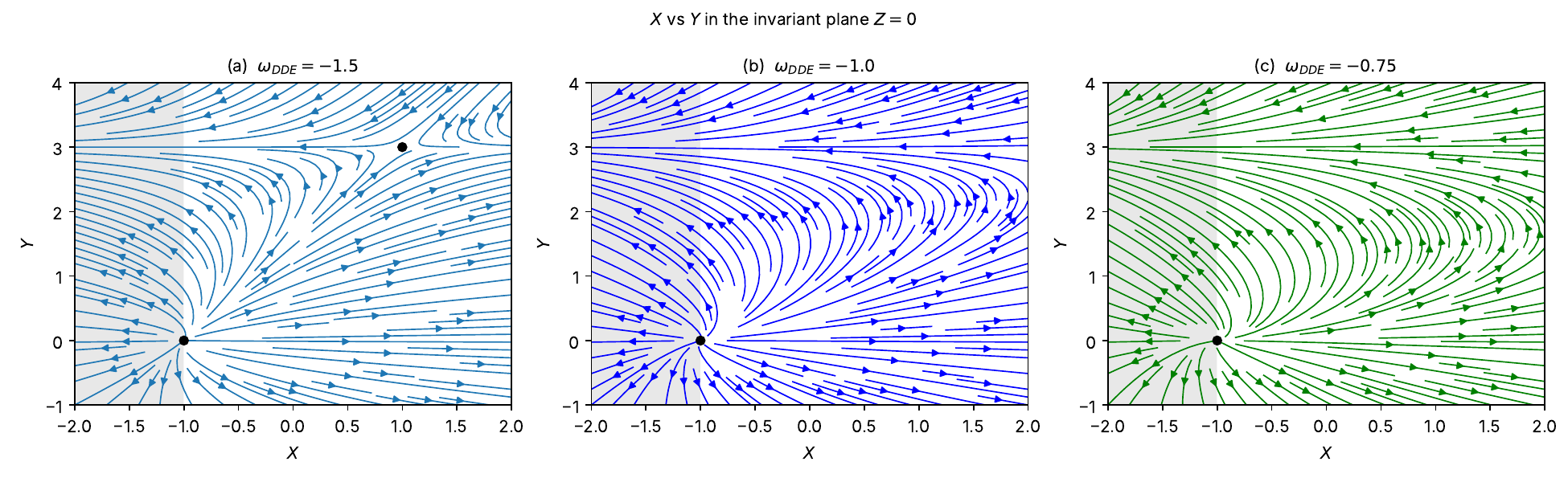}
\caption{Phase portraits $X$ (horizontal axis) and $Y$ (interaction) (vertical axis) in the invariant plane $Z=0$. (a) $\wD=-1.5$, (b) $\wD=-1.0$, (c) $\wD=-0.75$. Dots: $P_{dS}$ and, for $\wD=-1.5$, $P_Q$ at $(X_*,3)=(1,3)$; for $\wD=-0.75$, $P_Q$ lies at $X_*=-5$, outside the window.}
\label{fig:XY}
\end{figure}

\begin{figure}[t]
\centering
\includegraphics[width=\textwidth]{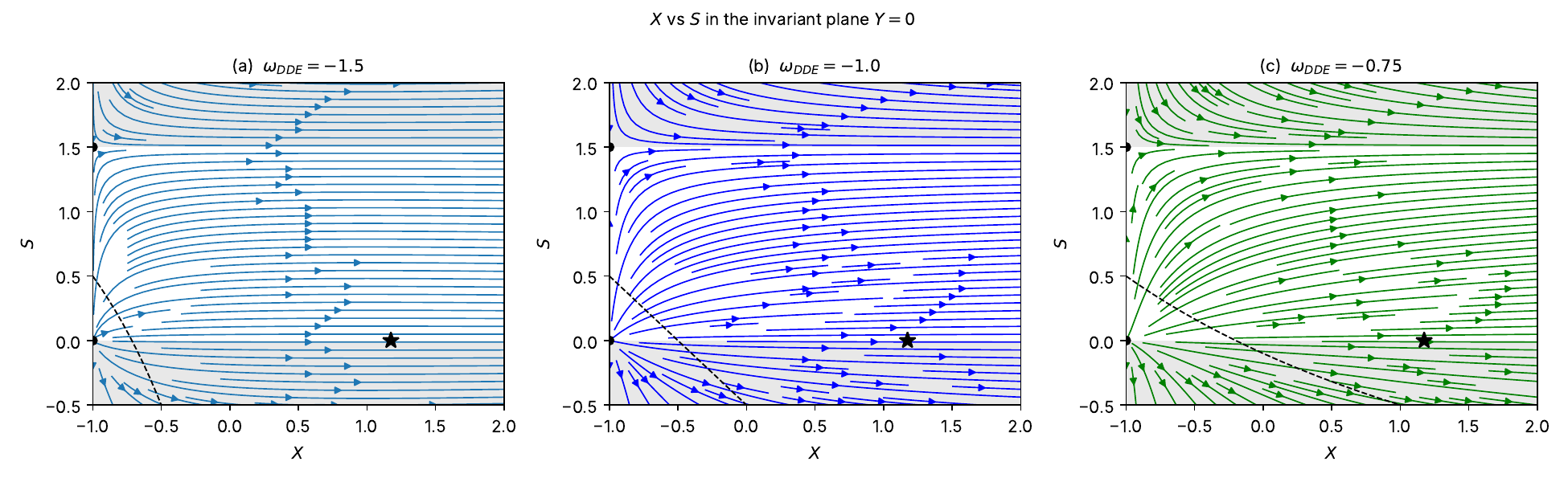}
\caption{Phase portraits $X$ (horizontal axis) and $S$ (cosmological constant) (vertical axis) in the invariant plane $Y=0$. (a) $\wD=-1.5$, (b) $\wD=-1.0$, (c) $\wD=-0.75$.}
\label{fig:XS}
\end{figure}

\begin{figure}[t]
\centering
\includegraphics[width=\textwidth]{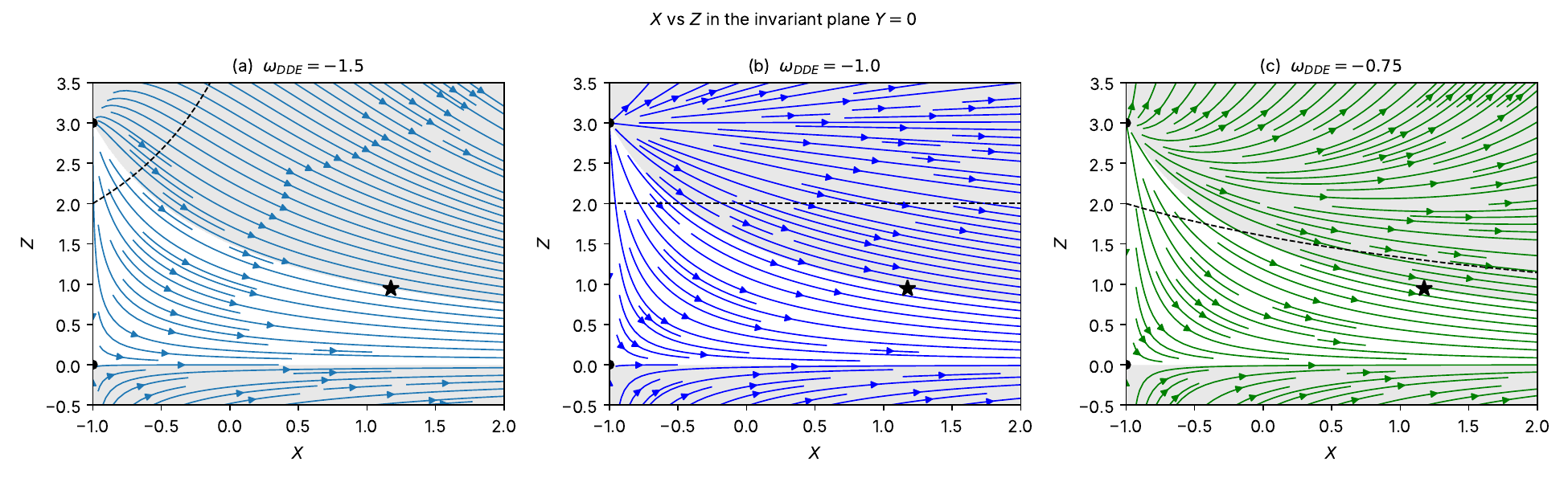}
\caption{Phase portraits $X$ (horizontal axis) and $Z$ (matter) (vertical axis) in the invariant plane $Y=0$. (a) $\wD=-1.5$, (b) $\wD=-1.0$, (c) $\wD=-0.75$. Dashed: acceleration boundary $Z=2/D$.}
\label{fig:XZ}
\end{figure}

\begin{figure}[t]
\centering
\includegraphics[width=\textwidth]{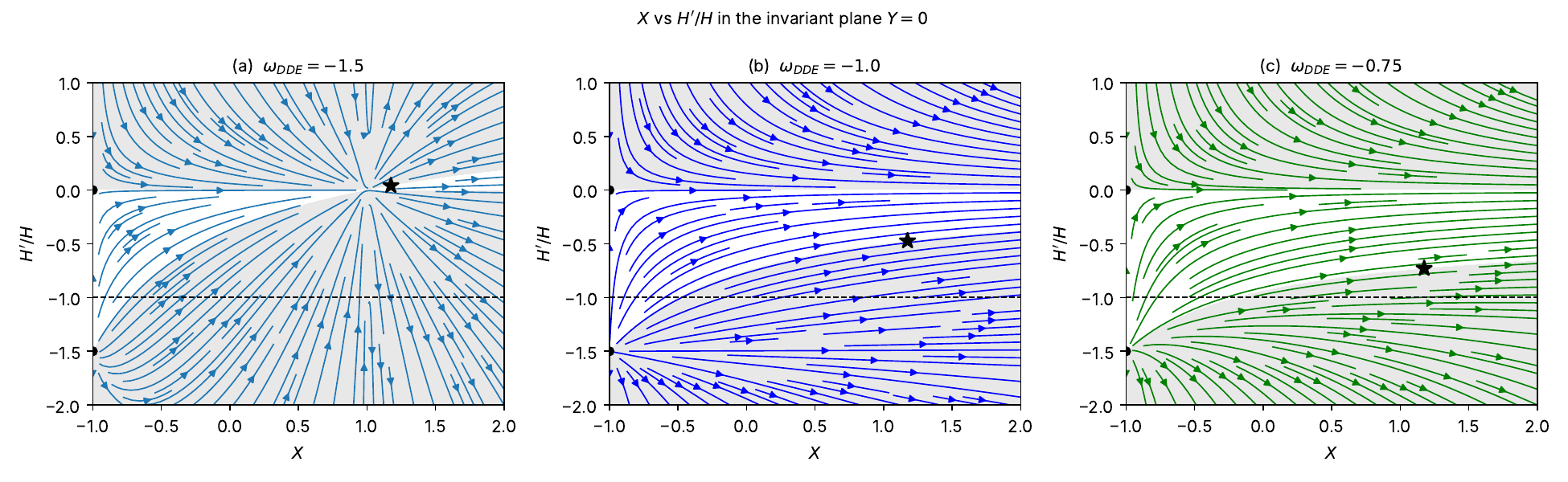}
\caption{As Fig.~\ref{fig:XH}, on the window $-1\leq X\leq2$, $-2\leq H'/H\leq1$; the pinch at $(1,0)$ in (a) is the same coordinate artefact.}
\label{fig:XHprime}
\end{figure}

These features are shown in Figs.~\ref{fig:XH}--\ref{fig:XHprime}. Each portrait is drawn in an invariant plane ($Y=0$ for Figs.~\ref{fig:XH}, \ref{fig:XS}, \ref{fig:XZ} and~\ref{fig:XHprime}; $Z=0$ for Fig.~\ref{fig:XY}), so it is an exact two-dimensional phase portrait rather than a slice. In Figs.~\ref{fig:XH} and~\ref{fig:XHprime} the plane $Y=0$ is mapped to $(X,H'/H)$ by Eq.~\eqref{eq:D}, and in Fig.~\ref{fig:XS} to $(X,S)$ by Eq.~\eqref{eq:Sslave}. Grey shading marks the region excluded by $Z\geq0$, $S\geq0$ or $X\geq-1$. The dashed curve is the acceleration boundary $H'/H=-1$ ($q=0$); by Eq.~\eqref{eq:D} it is $Z=2/D$ in the $(X,Z)$ plane. The star marks the present state for $\Omega_M=0.315$~\cite{Planck2018} with all dark energy dynamic ($\beta=0$, see below). No initial conditions are needed to draw the portraits: the vector field is evaluated on a uniform grid covering each window.

A key consequence concerns the commonly assumed identification of the onset of acceleration with matter--dark energy equality, $X=0$. At $X=0$ the constraint~\eqref{eq:constraint} gives $Z=(3-2S)/2\leq3/2$ whenever $\Lambda\geq0$, and $D=2+\wD$, so that
\begin{equation}
\left.\frac{H'}{H}\right|_{X=0}=-\frac{(2+\wD)(3-2S)}{4}\;\geq\;-\frac{3(2+\wD)}{4}. \label{eq:accel_equality}
\end{equation}
for $\wD>-2$ (for $\wD\leq-2$, $H'/H\geq0$ at $X=0$). This exceeds $-1$, i.e.\ $q<0$, for every $\wD<-2/3$. Hence, for any constant $\wD<-2/3$ and non-negative $\Lambda$, and independently of the interaction, the universe is already accelerating when $\rD=\rM$: acceleration begins before equality. Equality is not the transition point; the transition is set by $\wD$ and by the composition of the dark energy.

\begin{figure}[t]
\centering
\includegraphics[width=\textwidth]{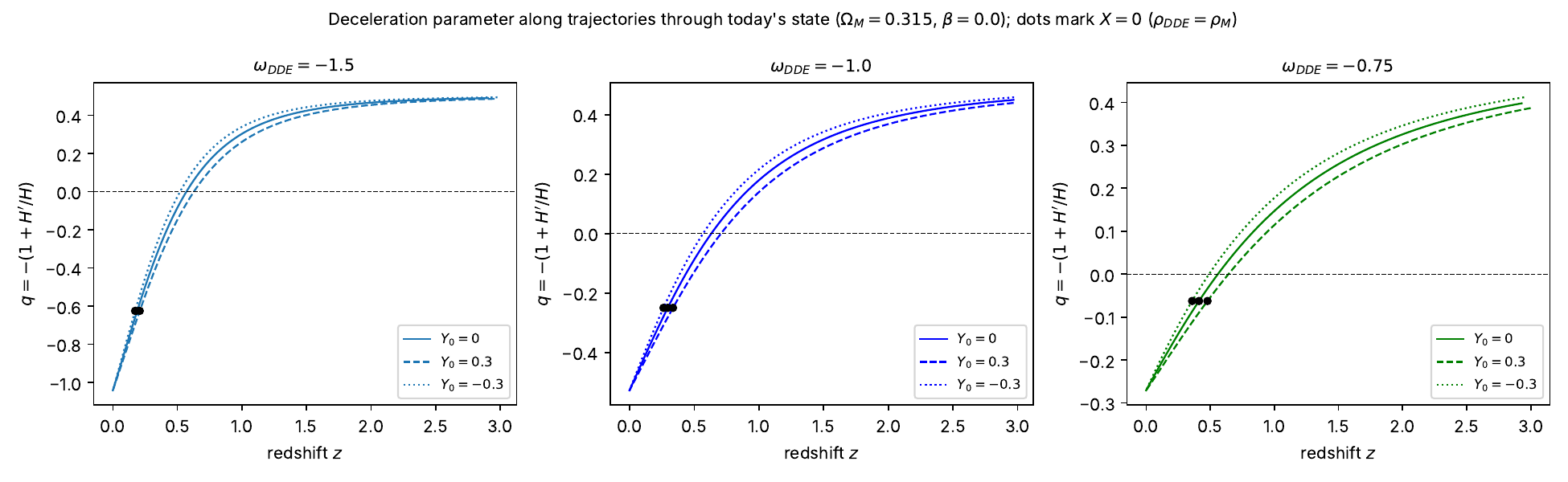}
\caption{Deceleration parameter $q=-(1+H'/H)$ against redshift along trajectories through the present state, Eq.~\eqref{eq:X0} with $\Omega_M=0.315$, $\beta=0$, for $Y_0=0$ (solid), $0.3$ (dashed) and $-0.3$ (dotted). Dots mark equality, $X=0$; acceleration ($q<0$) begins at higher redshift in every case.}
\label{fig:traj}
\end{figure}

To quantify this, we integrate the system from present-day data. Writing the observed ratio of dark energy to matter as $r=(1-\Omega_M)/\Omega_M\approx2.2$ and $\beta=\Lambda/\rD$ (with $8\pi G=1$),
\begin{equation}
\frac{\rD}{H^2}+\frac{\Lambda}{H^2}\approx r\,\frac{\rM}{H^2},\qquad \rD(1+\beta)\approx r\rM, \label{eq:today}
\end{equation}
which holds at the present epoch irrespective of $\wD$, and gives
\begin{equation}
X_0=\frac{r}{1+\beta}-1,\qquad Z_0=3\Omega_M,\qquad S_0=\frac{3-Z_0(X_0+2)}{2}. \label{eq:X0}
\end{equation}
For instance, $\beta=0$ gives $X_0\approx1.2$; $\beta=1$ ($\Lambda=\rD$) gives $X_0\approx0.09$; and $\beta=r-1\approx1.2$ gives $X_0=0$, so that if $\Lambda$ exceeds $\rD$ by more than this factor, the dynamic component is still subdominant to matter today. The discrepancy between the values of $\Lambda$ suggested by quantum field theory and by cosmology~\cite{Hobson2014} leaves $\beta$ open. Taking Eq.~\eqref{eq:X0} with $\beta=0$ as initial data at $\eta=0$ and integrating backwards gives Fig.~\ref{fig:traj} and Table~\ref{tab:traj}. In every case, the deceleration-to-acceleration transition occurs at $z_{\rm acc}\approx0.5$--$0.7$, while equality occurs later, at $z_{\rm eq}\approx0.2$--$0.5$. A non-zero interaction shifts both, $Y_0>0$ to earlier and $Y_0<0$ to later times, but does not reverse their order. The present deceleration parameter, $q_0=-1+\tfrac32[\Omega_M+(1+\wD)\Omega_{DDE}]$ for $Y_0=0$, is $q_0\approx-0.53$ for $\wD=-1$, in agreement with $q_0\approx-0.55$~\cite{Turner2000}, while $\wD=-1.5$ gives $q_0\approx-1.04$, within the range $q_0\approx-1.08\pm0.29$ inferred from local supernovae~\cite{Camarena2020}. Improved determinations of $H_0$ and $q_0$ would therefore constrain $X_0$, and through it the split $\beta$ of the dark energy.

\begin{table}[t]
\caption{Trajectories through the present state, Eq.~\eqref{eq:X0} with $\Omega_M=0.315$ and $\beta=0$: present deceleration parameter $q_0$, redshift of the onset of acceleration $z_{\rm acc}$ ($q=0$) and redshift of equality $z_{\rm eq}$ ($X=0$).}
\label{tab:traj}
\begin{ruledtabular}
\begin{tabular}{rrccc}
$\wD$ & $Y_0$ & $q_0$ & $z_{\rm acc}$ & $z_{\rm eq}$\\
\hline
$-1.5$ & $0$ & $-1.04$ & $0.57$ & $0.19$\\
$-1.5$ & $\pm0.3$ & $-1.04$ & $0.63\,/\,0.52$ & $0.21\,/\,0.17$\\
$-1.0$ & $0$ & $-0.53$ & $0.63$ & $0.30$\\
$-1.0$ & $\pm0.3$ & $-0.53$ & $0.71\,/\,0.57$ & $0.34\,/\,0.26$\\
$-0.75$ & $0$ & $-0.27$ & $0.56$ & $0.41$\\
$-0.75$ & $\pm0.3$ & $-0.27$ & $0.64\,/\,0.49$ & $0.48\,/\,0.36$\\
\end{tabular}
\end{ruledtabular}
\end{table}

Finally, $\wD$ acts as a control parameter for the flow. At $\wD=-1$ the line $\mathcal{L}$ leaves the phase space ($X_*\to\pm\infty$) and the eigenvalue $-3(1+\wD)$ of $P_Q$ changes sign; at $\wD=0$ the eigenvalue $-3\wD$ of $P_M$ and $P_{dS}$ changes sign. These are bifurcations of the phase portrait, visible in the change of the flow between panels (a) and (c) of each figure.

\section{Stability analysis}\label{sec:stability}

A fixed point is asymptotically stable if all trajectories that start close to it approach it as $\eta\to\infty$; by Section~3 this is determined, for hyperbolic points, by the signs of the real parts of the eigenvalues of the Jacobian. For the reduced system write
\begin{align}
f &= -3\wD(X+1)-Y(X+2), \nonumber\\
g &= Y\left(3-Y+\frac{ZD}{2}\right), \nonumber\\
h &= Z\left(-3+Y+ZD\right), \label{eq:fgh}
\end{align}
with $D$ given by Eq.~\eqref{eq:D}, so that $\partial D/\partial X=1+\wD$. The derivatives are
\begin{align}
f_X &= -3\wD-Y, & f_Y &= -(X+2), & f_Z &= 0, \label{eq:dfs}\\
g_X &= \frac{(1+\wD)YZ}{2}, & g_Y &= 3-2Y+\frac{ZD}{2}, & g_Z &= \frac{YD}{2}, \label{eq:dgs}\\
h_X &= (1+\wD)Z^2, & h_Y &= Z, & h_Z &= -3+Y+2ZD. \label{eq:dhs}
\end{align}
At the de~Sitter point $(-1,0,0)$,
\begin{equation}
J=\begin{pmatrix}
-3\wD & -1 & 0\\
0 & 3 & 0\\
0 & 0 & -3
\end{pmatrix},\qquad |\lambda I-J|=0\ \Rightarrow\ \lambda=-3\wD,\ 3,\ -3, \label{eq:J}
\end{equation}
a saddle for every $\wD\neq0$. The same calculation gives the eigenvalues listed in Table~\ref{tab:stab}. Along $\mathcal{L}$ the non-zero eigenvalues are
\begin{equation}
\lambda_\pm=-\frac{3(\wD+2)}{2}\pm\frac{\sqrt3}{2}\sqrt{-\wD\left(2Z-3\wD\right)}. \label{eq:eigL}
\end{equation}

\begin{table}[t]
\caption{Fixed points of the reduced system and their eigenvalues; ``nh'' denotes non-hyperbolic.}
\label{tab:stab}
\begin{ruledtabular}
\begin{tabular}{lccp{0.36\textwidth}}
Point & $(X,Y,Z)$ & Eigenvalues & Character ($\wD<0$)\\
\hline
$P_M$ & $(-1,0,3)$ & $3,\ \tfrac92,\ -3\wD$ & source\\
$P_{dS}$ & $(-1,0,0)$ & $3,\ -3,\ -3\wD$ & saddle\\
$P_Q$ & $(X_*,3,0)$ & $0,\ -3,\ -3(1+\wD)$ & nh; attracting in $Z=0$ if $\wD>-1$\\
$\mathcal{L}$ & $(X_*,3,Z)$ & $0,\ \lambda_+,\ \lambda_-$ & nh; saddle type if $\wD<-1$; attracting for $Z<-6(1+\wD)/\wD$ if $-1<\wD<0$\\
\end{tabular}
\end{ruledtabular}
\end{table}

\section{Conclusion}

In conclusion, our study has explored the intricate dynamics surrounding the transition between matter and dark energy, specifically considering scenarios with interacting constituents. Building on our exploration of matter-dark energy interactions, we employed dynamical systems techniques to delve into the critical factors influencing these dynamics. This approach enabled us to unpack the complex interplay between various cosmological constituents and shed light on their evolving characteristics as the universe transitions from a matter-dominated phase to a dark energy-dominated one. By understanding the role of key parameters and their influence over time, we can enhance our theoretical models and better align them with observational evidence. These insights are crucial for advancing our understanding of the universe's expansion history and future trajectory, illustrating the value of integrating dynamical systems methodologies in cosmological research.

Our analysis shows that the matter-dominated state is the past attractor of the flow and that, without interaction, the late-time evolution in the variable $X$ is towards dynamic dark energy domination. A constant interaction does not survive to late times: for $Q_{ME}<0$ the matter density vanishes at finite scale factor, and for $Q_{ME}>0$ with $\wD\geq-1$ the dynamic dark energy density becomes negative. The analysis also shows that the onset of acceleration does not coincide with matter--DDE equality: for any constant $\wD<-2/3$ and non-negative $\Lambda$ the universe is already accelerating when $\rD=\rM$, and with present-day data the transition occurs at $z\approx0.5$--$0.7$, against $z\approx0.2$--$0.5$ for equality. The equation of state $\wD$ is a control parameter whose variation produces bifurcations at $\wD=-1$ and $\wD=0$. The inferred transition epoch therefore depends on $\wD$ and on how the dark energy is split between its dynamic and non-dynamic parts, which suggests that values of $\wD$ reconstructed from observations should be interpreted with the epoch of the data in mind. Natural extensions are an evolving $\wD$ and interactions of the form $Q_{ME}\propto H\rho$, which avoid the late-time pathologies of a constant coupling.

\begin{acknowledgments}
The author thanks the University of Cape Town's NGP for financial support.
\end{acknowledgments}


\end{document}